# Organized Self-Emulsification toward Structural Color

Xi Chen,[1] Xiao Yang,[2] Dong-Po Song,[1,*] Yuesheng Li[1]

[1] Tianjin Key Laboratory of Composite and Functional Materials, School of Materials Science and Engineering, Tianjin University, Tianjin 300350, China.

[2] Changchun Institute of Applied Chemistry, Chinese Academy of Sciences, Changchun 130022, China.

*Supporting Information Placeholder*

**ABSTRACT:** The formation of water-in-oil-in-water (W/O/W) double emulsions can be well-controlled through an organized self-emulsification mechanism in the presence of rigid bottlebrush amphiphilic block copolymers. Nanoscale water droplets with well-controlled diameters form ordered spatial arrangements within the micron-scale oil droplets. Upon solvent evaporation, solid microspheres with hexagonal close packed nanopore arrays are obtained resulting in bright structural colors. The reflected color is precisely tunable across the whole visible light range through tailoring contour length of the bottlebrush molecule. In-situ observation of the W/O interface using confocal laser scanning microscopy provides insights into the mechanism of the organized self-emulsification. This work provides a powerful strategy for the fabrication of structural colored materials in an easy and scalable manner.

Nature created vivid colors via light absorption by pigments and/or reflection from periodic nanostructures, and the latter case is known as structural color. In comparison with pigmental color, structural color exhibits many unique features such as the resistance to photo/chemical bleaching, their stimuli responsive behaviors, and reduced toxicity.[1] Structural color is typically created via creation of liquid crystalline phase, microphase separation of block copolymers (BCPs), or colloidal crystallization.[2-3] Self-assembly of colloidal particles is the most widely used method to create three dimensional (3D) photonic crystals including opal or inverse opal structures. In particular, inverse opal is of great interest for optics, sensing and energy applications due to their ordered porous structure with large specific surface areas.[4] However, such porous structure typically requires time-consuming multi-steps for the fabrication process and harsh conditions for subsequent etching to generate pores inside a polymer matrix such as using hydrofluoric acid, which brings a significant barrier to large-scale applications.

In this work, we show that 3D ordered porous microspheres can be readily fabricated via one-step formation of ordered water-in-oil-in-water (W/O/W) double emulsions in the presence of amphiphilic bottlebrush block copolymers (BBCPs). (Polynorbornene-g-polystyrene)-b-(polynorbornene-g-polyethylene oxide) (PS-b-PEO) BBCPs were selected to demonstrate the facile fabrication of structural colored materials via a novel organized self-emulsification mechanism (see Figure 1). The molecular characteristics and chemical structure characteristics are summarized in Table S1, Figure S1 and S2 (Supporting Information). It should be noted that the BBCPs with a low degree of polymerization (53 < DP < 77 repeat units) are employed for creating structural color. This is different from previous works in which self-assembly of BBCPs with huge molecular weights (MWs) are typically required for making nanostructures large enough to reflect visible light.[5-12] Synthesis of BCPs with very large MWs is challenging, and the obtained samples may have solubility issue when subjected to solution processing.

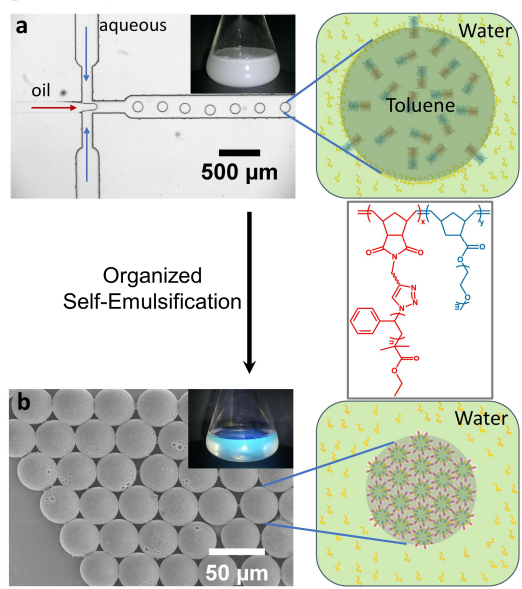

**Figure 1.** (a) Preparation of oil in water emulsion using a microfluidic device. The inset is a photograph of the obtained emulsion. (b) SEM micrograph of the porous microspheres after solvent evaporation. The inset is a photograph of the obtained microspheres showing a bright structural color.

Figure 1a shows the simple fabrication procedure involved in this work. BBCPs are dissolved in toluene forming clear solutions followed by emulsification in water containing polyvinyl alcohol (PVA) as the stabilizer (see Supporting Information for details). A flow-focusing microfluidic device was used to generate monodisperse microdroplets with a diameter of 120 μm leading to solid microspheres of approximately 35 μm upon drying of the organic solvent as measured based on field emission scanning microcopy (SEM) (Figure 1b). W/O/W double emulsions are generated during the evaporation of the organic solvent, and the synergistic adsorption of bottlebrush amphiphiles and PVA onto O/W interface can induce a significant decrease in interfacial tension (Figure S3) leading to interfacial instability. Consequently, interface roughening happened with small water droplets infiltrated the toluene phase forming the W/O/W double emulsions. Such phenomenon has been reported with linear BCPs as the amphiphiles, and various interesting structures were created such as spherical micelle, worm-like micelles, foam-like porous particles etc.[13-15] However, well-control over the self-emulsification process to create ordered periodic nanostructures showing bright structural colors has proven elusive.

We demonstrate that the bottlebrush amphiphiles, rigid surfactants, can produce nano-scale water droplets with well-controlled diameters within the toluene phase. Subsequently, ordered spatial arrangement of the water droplets are achieved resulting in the 3D ordered porous structure with bright structural colors upon drying of the organic solvent (Figure 1b and Figure 2). For contrast, hierarchical porous structures were typically obtained with no structural color when using a linear PS-b-PEO BCP.[14b] The highly-extended molecular conformation of the bottlebrush polymer turns out to be the key factor that enables such organized self-emulsification mechanism (see below).

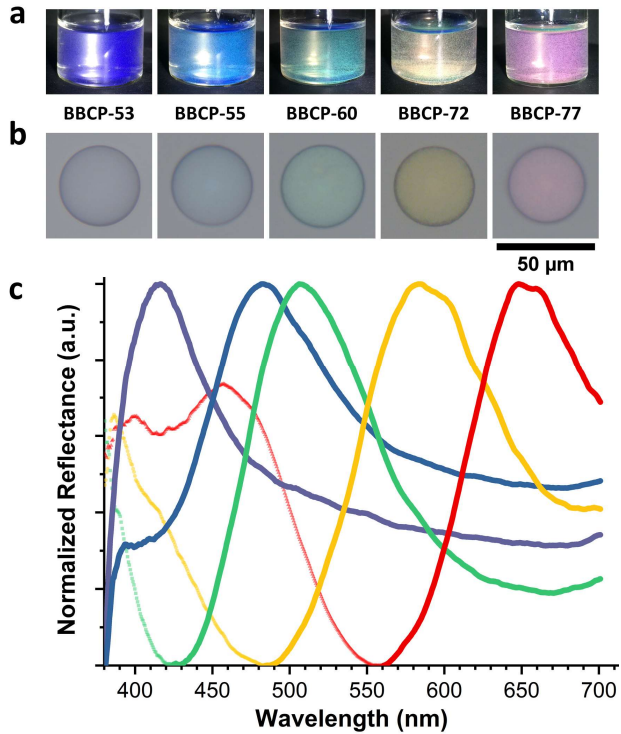

**Figure 2.** (a) Photographs and (b) optical micrographs (reflection mode) of the microspheres dispersed in aqueous phase showing varied structural colors dependent on DP. (c) Reflection spectra of the microspheres within the aqueous phase.

We keep the volume fraction of PEO constant (~50%) to investigate the influence of total degree of polymerization (DP) on the porous structure and structural color. Here BBCPs with different DPs are labeled as BBCP-DP for the following statements. The reflected color is precisely tunable via changing the backbone length of the bottlebrush molecules (DP). Violet, blue, green, yellow and red colors are obtained using bottlebrushes with total DPs of approximately 53, 55, 60, 72 and 77, respectively (Figure 2a). Further increase DP to 88 results in a sample exhibiting a reflected color in near infrared range. Optical microscopy on individual microspheres in water revealed bright color reflections (Figure 2b and Figure S4) with evident peaks in the reflection spectra (Figure 2c). This is consistent with their optical appearance on a macroscopic scale (Figure 2a). The intense reflection suggests the formation of an ordered periodic structure inside the microspheres. The reflection maximum and their shift as a function of structural parameters can be well described by using the following equation:

$$\lambda = 2d\sqrt{n_{avg}^2 - n_{void}^2 \sin^2\theta} \tag{1}$$

where $\lambda$ is the wavelength of reflection maximum, $d$ is the spacing between closed packed planes of pores, $n_{avg}$ is the average refractive index (RI) of the materials, $n_{void}$ is the RI of aqueous phase in the void, and $\theta$ is the angle with respect to the normal to the close-packed planes.[4a] According to equation 1, angle-dependent iridescence was also observed for the obtained samples indicating a well-ordered structure obtained (Figure S5 and Figure 3). In addition, a blue shift of the reflection peak was observed for the reflection spectra of dried samples with the pores filled with air relative to that of the samples in water due to both decreased $n_{avg}$ and lattice spacing ($d$) caused by a small shrinkage of the structure (Figure S6).

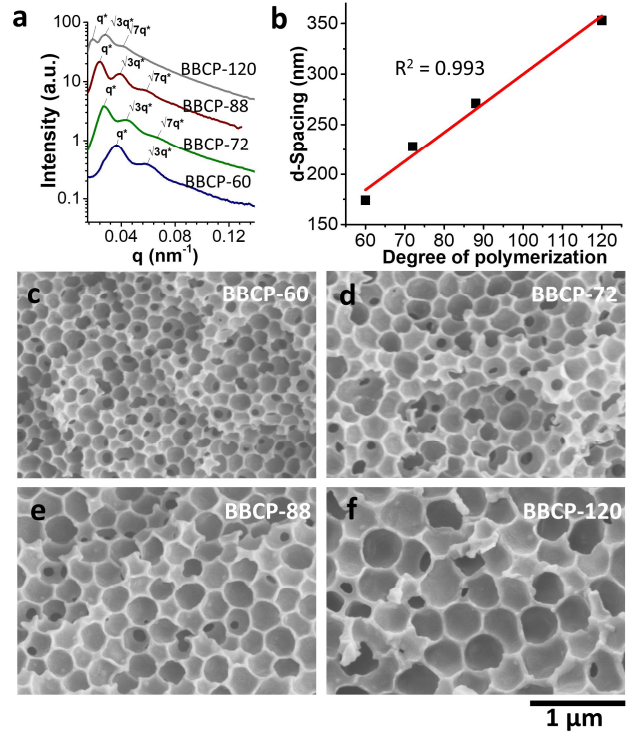

**Figure 3.** (a) USAXS spectra of the microspheres prepared using BBCPs with different DP. (b) Average domain spacing of the periodic porous structure (USAXS) as a function of DP. (c-f) SEM micrographs showing the cross-sections of different microspheres with different pore diameters of 181±18, 260±15, 299±13 and 395±19 nm, respectively.

Ultra-small angle X-ray scattering (USAXS) and cross-sectional SEM are employed to characterize the porous structure formed within the dried microspheres (Figure 3). Three ordering peaks are evident in the USAXS spectra (Figure 3a) for the samples prepared using BBCPs with DPs varying from 60 to 120 indicating long-range-ordered periodic structures formed. The $q$ ratio of the three peaks is close to 1: √3: √7 corresponding to hexagonal close packed (h.c.p.) nanopore arrays.[16] This is consistent with their angle-dependent color change arising from the well-ordered structure (Figure S5). The average d-spacing of the periodic structure increases linearly from 174 to 353 nm with the increase of DP from 60 to 120 as calculated according to $d = 2\pi/q^*$ (Figure 3b) which is close to the values measured based on cross-sectional SEM (Figure 3c-f). The increase of d-spacing results in a red shift of the reflected color according to equation 1 consistent with the optical behaviors observed (Figure 2). The well-control of the pore size is achieved via an organized self-emulsification mechanism enabled by the rigid bottlebrush amphiphiles (see below).

The organized self-emulsification mechanism was disclosed via in-situ monitoring the evolution of W/O interface during solvent evaporation using confocal laser scanning microscopy. For convenient observation, the W/O interface is created by adding a few drops of BBCP-120 solution in toluene onto the surface of PVA solution in water. A fluorescent dye was dissolved in the toluene phase to create a sharp contrast between oil and water phases. In our case, the dark and bright regions represent water and oil phases, respectively. Figure 4a shows the fluorescent micrographs of the W/O interface as a function of evaporation time. Water droplets with a broad size distribution and a few micrometer-large droplets are formed simultaneously upon creation of the W/O interface, which is similar to that observed previously when using linear BCPs.[14] More large water droplets and better contrast are confirmed after 50 minutes of solvent evaporation, indicating more water was transferred into the oil phase. However, no evident control of both the droplet size and spatial arrangement was achieved at this point probably because the adsorption of bottlebrush amphiphiles onto W/O interface is not dense enough to control droplet size and create an ordered state. A significant change was observed after 100 minutes of evaporation, and the large water droplets were clearly splitted into nearly uniform nanodroplets as highlighted in Figure 4a suggesting more BBCP molecules are involved in forming the inner water droplets (see Figure 4b). Upon evaporation for 165 minutes, the oil phase is nearly dried and ordered arrangements of the water nanodroplets is achieved. Concentrated suspension of nearly uniform water droplets crystallizes driven by entropy which is similar to that of colloids crystallization: The free volume of hard spheres is greater than that of disordered state resulting in higher entropy in the ordered state.[17] SEM micrograph (Figure 4c) of the dried film shows ordered spatial arrangement of the obtained nanopores, which is similar to the structure observed in the microspheres. The ordered porous structure was further confirmed by the bright blue color observed as well as the SEM micrograph for the dried film using BBCP-60 (Figure 4d). Moreover, organized self-emulsification was also achieved without PVA in the aqueous phase affording a dried film with an ordered structure and a bright reflected color (see Figure S7). This suggests that BBCP amphiphile is playing a critical role in the organized self-emulsification mechanism with limited contribution from PVA. The stable W/O emulsions obtained indicates the high performance of the BBCP surfactants, which is similar to previous studies on W/O or O/W emulsions using Janus bottlebrush amphiphiles as the effective stabilizer.[18]

Rapid self-assembly behavior of BBCPs has been reported in bulk benefiting from much less polymer chain entanglement relative to that of linear analogues.[19] Analogously, the highly extended molecular conformation of the bottlebrush can not only enable rapid adsorption and desorption of the bottlebrush molecules at W/O interface to approach a hydrodynamic equilibrium state but also form a thin film with the unique molecular arrangement on the W/O interface resulting in water droplets with well-controlled spherical curvature. For comparison, it is much more difficult to achieve the well-control using flexible linear BCPs as kinetically trapped state typically exists and the flexible molecules can form a variety of different molecular packing manners. In order to further confirm that the organized self-emulsification is greatly benefiting from the rigid conformation of the BBCP, we synthesized a triblock copolymer (Polynorbornene-g-polystyrene)-b-(polynorbornene-6-bromohexanoate)-b-(polynorbornene-g-polyethylene oxide) (PS-b-NBH-b-PEO) with a linear polymer block as the flexible spacer which will greatly decrease the rigidity of the molecular backbone. As a result, a broad reflection peak was observed for the microspheres as shown in the reflection spectrum, corresponding to a porous structure with less control on pore size as well as their spatial arrangement as indicated by cross-sectional SEM (See Figure S8 for details). The influence and theoretical study of more molecular parameters on the obtained nanostructure is a subject of future studies.

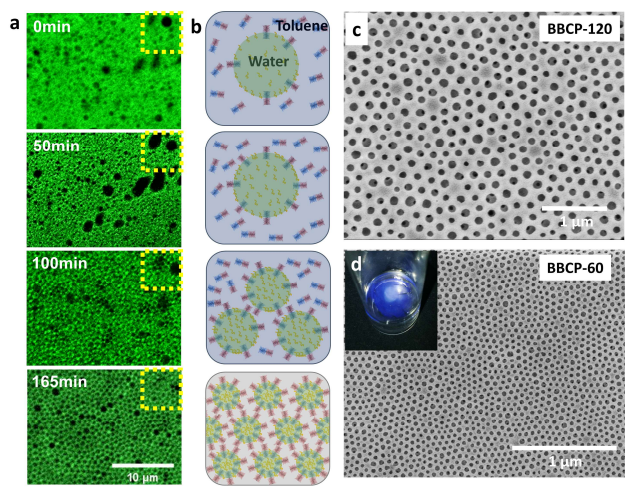

**Figure 4.** (a) Fluorescent micrographs of the W/O interface using confocal laser scanning microscopy. (b) Illustration of the large droplet splitting into uniform small droplets with ordered spatial arrangement. (c) SEM micrograph of the dried film using BBCP-120. (d) Photograph and SEM micrograph of a dried film prepared using BBCP-60.

In summary, we have demonstrated a powerful strategy for facile fabrication of ordered porous materials with bright structural colors via forming ordered W/O/W double emulsions. The reflected color is tunable across the whole visible light range greatly dependent on DP of the BBCP. An organized self-emulsification mechanism enabled by bottlebrush amphiphiles was disclosed via in-situ observation of the W/O interface evolution. In addition, BBCPs with only small DPs are required to generate structural color which makes it easy for the chemical synthesis. The one-step fabrication process has great potential for large-scale fabrication of photonic materials at low cost.

## ASSOCIATED CONTENT

### Supporting Information

Experimental information, supplementary figures. This material is available free of charge via the Internet at http://pubs.acs.org.

## AUTHOR INFORMATION

### Corresponding Author
*dongpo.song@tju.edu.cn

### Notes
The authors declare no competing financial interests.

## ACKNOWLEDGMENT

This work was supported by the National Natural Science Foundation of China [Grant No. 51873098]. We thank Prof. Yongfeng Men of Changchun Institute of Applied Chemistry for his assistance in USAXS measurements, Prof. Ziyi Yu of Nanjing Tech University for his help with microfluidics, and Prof. Silvia Vignolini of Cambridge University for valuable discussions.

# REFERENCES

(1) (a) Fu, Y.; Tippets, C. A.; Donev, E. U.; Lopez, R. Structural colors: from natural to artificial systems. *Wiley Interdiscip. Rev. Nanomedicine Nanobiotechnology* **2016**, *8*, 758-775. (b) Johansen, V. E.; Onelli, O. D.; Steiner, L. M.; Vignolini, S. Photonics in nature: from order to disorder. *Functional Surfaces in Biology III: Diversity of the Physical Phenomena* **2017**, *10*, 53-89. (c) Moyroud, E.; Wenzel, T.; Middleton, R.; Rudall, P. J.; Banks, H.; Reed, A.; Mellers, G.; Killoran, P.; Westwood, M. M.; Steiner, U.; Vignolini, S.; Glover, B. J. Disorder in convergent floral nanostructures enhances signalling to bees. *Nature* **2017**, *550*, 469-474.

(2) (a) Lee, S. S.; Kim, S. K.; Won, J. C.; Kim, Y. H.; Kim, S. H. Reconfigurable photonic capsules containing cholesteric liquid crystals with planar alignment. *Angew. Chem., Int. Ed.* **2015**, *54*, 15266-15270. (b) Lee, J.-H.; Koh, C. Y.; Singer, J. P.; Jeon, S.-J.; Maldovan, M.; Stein, O.; Thomas, E. L. Ordered polymer structures for the engineering of photons and phonons. *Adv. Mater.* **2014**, *26*, 532-569. (c) Stefik, M.; Guldin, S.; Vignolini, S.; Wiesner, U.; Steiner, U. Block copolymer self-assembly for nanophotonics. *Chem. Soc. Rev.* **2015**, *44*, 5076-5091. (d) Zhao, Y.; Shang, L.; Cheng, Y.; Gu, Z. Spherical colloidal photonic crystals. *Acc. Chem. Res.* **2014**, *47*, 3632-3642. (e) Wu, P.; Wang, J.; Jiang, L. Bio-inspired photonic crystal patterns. *Mater. Horiz.* **2020**, *7*, 338-365.

(3) (a) Goerlitzer, E. S. A.; Klupp Taylor, R. N.; Vogel, N. Bioinspired photonic pigments from colloidal self-assembly. *Adv. Mater.* **2018**, *30*, 1706654. (b) Vogel, N.; Utech, S.; England, G. T.; Shirman, T.; Phillips, K. R.; Koay, N.; Burgess, I. B.; Kolle, M.; Weitz, D. A.; Aizenberg, J. Color from hierarchy: diverse optical properties of micron-sized spherical colloidal assemblies. *Proc. Natl. Acad. Sci. U. S. A.* **2015**, *112*, 10845-10850. (c) Xiao, M.; Hu, Z.; Wang, Z.; Li, Y.; Tormo, A. D.; Thomas, N. L.; Wang, B.; Gianneschi, N. C.; Shawkey, M. D.; Dhinojwala, A.; Bioinspired bright noniridescent photonic melanin supraballs. *Sci. Adv.* **2017**, *3*, e1701151. (d) Zhang, J.; Meng, Z.; Liu, J.; Chen, S.; Yu, Z. Spherical Colloidal Photonic Crystals with Selected Lattice Plane Exposure and Enhanced Color Saturation for Dynamic Optical Displays. *ACS Appl. Mater. Interfaces* **2019**, *11*, 42629-42634. (e) Wu, X.; Hong, R.; Meng, J.; Cheng, R.; Zhu, Z.; Wu, G.; Li, Q.; Wang, C.-F.; Chen, S. Hydrophobic Poly(tert-butyl acrylate) Photonic Crystals towards Robust Energy-Saving Performance. *Angew. Chem. Int. Ed.* **2019**, *58*, 13556-13564.

(4) (a) Aguirre, C. I.; Reguera, E.; Stein A. Tunable colors in opals and inverse opal photonic crystals. *Adv. Funct. Mater.* **2010**, *20*, 2565-2578. (b) Armstrong, E.; O'Dwyer, C. Artificial opal photonic crystals and inverse opal structures-fundamentals and applications from optics to energy storage. *J. Mater. Chem. C* **2015**, *3*, 6109-6143. (c) Couturier, J.-P.; Sütterlin, M.; Laschewsky, A.; Hettrich, C.; Wischerhoff, E. Responsive inverse opal hydrogels for the sensing of macromolecules. *Angew. Chem., Int. Ed.* **2015**, *54*, 6641-6644. (d) Zhu, B.; Fu, Q.; Chen, K.; Ge, J. Liquid photonic crystals for mesopore detection. *Angew. Chem., Int. Ed.* **2018**, *57*, 252-256.

(5) (a) Rzayev, J. Molecular bottlebrushes: new opportunities in nanomaterials fabrication. *ACS Macro Lett.* **2012**, *1*, 1146-1149. (b) Fenyves, R.; Schmutz, M.; Horner, I. J.; Bright, F. V.; Rzayev, J. Aqueous self-assembly of giant bottlebrush block copolymer surfactants as shape-tunable building blocks. *J. Am. Chem. Soc.* **2014**, *136*, 7762-7770. (c) Li, Z.; Ma, J.; Cheng, C.; Zhang, K.; Wooley, K. L. Synthesis of hetero-grafted amphiphilic diblock molecular brushes and their self-assembly in aqueous medium. *Macromolecules* **2010**, *43*, 1182-1184.

(6) Liberman-Martin, A. L.; Chu, C. K.; Grubbs, R. H. Application of bottlebrush block copolymers as photonic crystals. *Macromol. Rapid Commun.* **2017**, *38*, 1700058.

(7) Miyake, G. M.; Piunova, V. A.; Weitekamp, R. A.; Grubbs, R. H. Precisely tunable photonic crystals from rapidly self-assembling brush block copolymer blends. *Angew. Chem., Int. Ed.* **2012**, *51*, 11246-11248.

(8) Sveinbjörnsson, B. R.; Weitekamp, R. A.; Miyake, G. M.; Xia, Y.; Atwater, H. A.; Grubbs, R. H. Rapid Self-Assembly of brush block copolymers to photonic crystals. *Proc. Natl. Acad. Sci. U. S. A.* **2012**, *109*, 14332-14336.

(9) Macfarlane, R. J.; Kim, B.; Lee, B.; Weitekamp, R. A.; Bates, C. M.; Lee, S. F.; Chang, A. B.; Delaney, K. T.; Fredrickson, G. H.; Atwater, H. A.; Grubbs, R. H. Improving brush polymer infrared one-dimensional photonic crystals via linear polymer additives. *J. Am. Chem. Soc.* **2014**, *136*, 17374-17377.

(10) Song, D.-P.; Li, C.; Colella, N. S.; Lu, X.; Lee, J.-H.; Watkins, J. J. Thermally tunable metallodielectric photonic crystals from the self-assembly of brush block copolymers and gold nanoparticles. *Adv. Opt. Mater.* **2015**, *3*, 1169-1175.

(11) Song, D.-P.; Li, C.; Li, W.; Watkins, J. J. Block copolymer nanocomposites with high refractive index contrast for one-step photonics. *ACS Nano* **2016**, *10*, 1216-1223.

(12) Song, D.-P.; Zhao, T. H.; Guidetti, G.; Vignolini, S.; Parker, R. M. Hierarchical photonic pigments via the confined self-assembly of bottlebrush block copolymers. *ACS Nano* **2019**, *13*, 1764-1771.

(13) (a) Hanson, J. A.; Chang, C. B.; Graves, S. M.; Li, Z.; Mason, T. G.; Deming, T. J. Nanoscale double emulsions stabilized by single-component block copolypeptides. *Nature* **2008**, *455*, 85-88. (b) Hong, L.; Sun, G.; Cai, J.; Ngai, T. One-step formation of w/o/w multiple emulsions stabilized by single amphiphilic block copolymers. *Langmuir* **2012**, *28*, 2332-2336. (c) Besnard, L. ; Marchal, F.; Paredes, J. F.; Daillant, J.; Pantoustier, N.; Perrin, P.; Guenoun, P. Multiple emulsions controlled by stimuli-responsive polymers. *Adv. Mater.* **2013**, *25*, 2844-2848.

(14) (a) Zhu, J.; Hayward, R. C. Spontaneous generation of amphiphilic block copolymer micelles with multiple morphologies through interfacial instabilities. *J. Am. Chem. Soc.* **2008**, *130*, 7496-7502. (b) Zhu, J.; Hayward, R. C. Hierarchically structured microparticles formed by interfacial instabilities of emulsion droplets containing amphiphilic block copolymers. *Angew. Chem., Int. Ed.* **2008**, *47*, 2113-2116.

(15) (a) Bae, J.; Russell, T. P.; Hayward, R. C. Osmotically driven formation of double emulsions stabilized by amphiphilic block copolymers. *Angew. Chem., Int. Ed.* **2014**, *53*, 8240-8245. (b) Ku, K. H.; Shin, J. M.; Klinger, D.; Jang, S. G.; Hayward, R. C.; Hawker, C. J.; Kim, B. J. Particles with tunable porosity and morphology by controlling interfacial instability in block copolymer emulsions. *ACS Nano* **2016**, *10*, 5243-5251. (c) Shin, J. J.; Kim, E. J.; Ku, K. H.; Lee, Y. J.; Hawker, C. J.; Kim, B. J. 100th Anniversary of Macromolecular Science Viewpoint: Block Copolymer Particles: Tuning Shape, Interfaces, and Morphology. *ACS Macro Lett.* **2020**, *9*, 306-317.

(16) Zheng, N.; Yi, Z.; Li, Z.; Chen, R.; Lai, Y.; Men, Y. Achieving grazing-incidence ultra-small-angle X-ray scattering in a laboratory setup. *J. Appl. Cryst.* **2015**, *48*, 608-612.

(17) Gasser, U.; Weeks, E. R.; Schofield, A.; Pusey, P. N.; Weitz, D. A. Real-space imaging of nucleation and growth in colloidal crystallization. *Science* **2001**, *292*, 258-262.

(18) (a) Li, Y.; Zou, J.; Das, B. P.; Tsianou, M.; Cheng, C. Well-defined amphiphilic double-brush copolymers and their performance as emulsion surfactants. *Macromolecules* **2012**, *45*, 4623-4629. (b) Xie, G.; Krys, P.; Tilton, R. D.; Matyjaszewski, K. Heterografted molecular brushes as stabilizers for water-in-oil emulsions. *Macromolecules* **2017**, *50*, 2942-2950.

(19) Song, D.-P.; Li, C.; Colella, N. S.; Xie, W.; Li, S.; Lu, X.; Gido, S.; Lee, J.-H.; Watkins, J. J. Large volume self-organization of polymer/nanoparticle hybrids with millimeter scale grain sizes using brush block copolymers. *J. Am. Chem. Soc.* **2015**, *137*, 12510-12513.

Insert Table of Contents artwork here

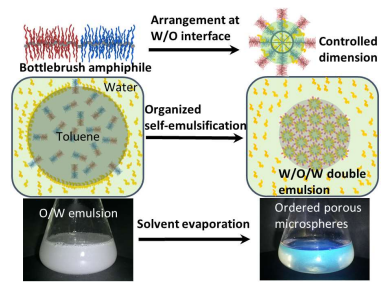